\documentstyle[aps,prl,twocolumn,epsf]{revtex}

\begin{document}
\twocolumn[\hsize\textwidth\columnwidth\hsize\csname
@twocolumnfalse\endcsname
\draft
\tightenlines

\title{Phase separation in star polymer-colloid mixtures} 

\author{J. Dzubiella,$^{1}$ A. Jusufi,$^{1}$ C. N. Likos,$^{1}$
        C. von Ferber,$^{1}$ H. L{\"o}wen,$^{1}$\\
        J. Stellbrink,$^{2,3}$
        J. Allgaier,$^{2}$ D. Richter,$^{2}$ A. B. Schofield,$^{3}$
        P. A. Smith,$^{3}$ W. C. K. Poon,$^{3}$ and P. N. Pusey$^{3}$} 
\address{$^{1}$ Institut f{\"u}r Theoretische Physik II,
         Heinrich-Heine-Universit\"at,
         Universit\"atsstra{\ss}e 1, D-40225 D\"usseldorf, Germany\\
         $^{2}$ IFF-Neutronenstreuung,
         Forschungszentrum J\"ulich GmbH, D-52425
         J\"ulich, Germany\\
         $^{3}$Department of Physics and Astronomy, The University of
         Edinburgh, Mayfield Road, Edinburgh EH9 3JZ, United Kingdom}

\date{Submitted to Phys.\ Rev.\ Lett., \today}

\maketitle
\begin{abstract}
We examine the demixing transition in star polymer-colloid
mixtures for star arm numbers $f=2,6,16,32$ and
different star-colloid size ratios. Theoretically, we solve the
thermodynamically self-consistent Rogers-Young integral equations for binary
mixtures using three effective pair potentials obtained from direct molecular
computer simulations. The numerical results show a spinodal instability.
The demixing binodals are approximately calculated, and 
found to be consistent with experimental observations. 
\end{abstract}
\vskip2pc]

\renewcommand{\thepage}{\hskip 8.9cm \arabic{page} \hfill Typeset
                        using REV\TeX }

\narrowtext

The study of mixtures of hard colloidal particles and
nonadsorbing polymer {\it chains} has received a great deal of recent
attention, both 
experimentally \cite{ilett:etal:pre:95,moussaid:etal:prl:99,lekkerkerker:etal:92} and
theoretically \cite{lekkerkerker:etal:92,dijkstra:brader:evans:jpcm:99,ard:finken:hansen:99,fuchs:schweizer:epl:00}. However, very little is known about
the phase behavior of the much richer mixture of colloidal particles
and {\it star} polymers, the latter being macromolecules
consisting of $f$ polymeric chains 
covalently attached on a common center. 
The theoretical approaches to the study of colloid-polymer mixtures
were largely based on the Asakura-Oosawa (AO) model, in which the chains
are envisaged as noninteracting spheres experiencing a hard-sphere
repulsion with the colloids. Subsequently, the system can be mapped
onto an effective one-component fluid featuring the so-called
depletion interaction between the colloids, mediated by the
ideal chains \cite{lekkerkerker:etal:92,dijkstra:brader:evans:jpcm:99}.
Though the AO model provides an excellent benchmark for such systems,
recent theoretical studies \cite{ard:finken:hansen:99,fuchs:schweizer:epl:00} 
and comparisons with experiments \cite{moussaid:etal:prl:99} indicate 
that the assumption of noninteracting chains leads to quantitative
discrepancies between the two. Hence, a systematic effort to
derive more realistic 
chain-chain \cite{ard:peter:00} as well
as chain-colloid \cite{fuchs:schweizer:epl:00,bolhuis:preprint:00} interactions
has already been undertaken. 

For the case of a star polymer-colloid mixture, the need to employ realistic
interactions is even more apparent.  Star polymers with a large functionality
$f$ are stiff particles \cite{likos:etal:prl:98,jusufi:macromolecules:99}  and
their mutual interactions cannot be  ignored, even as a first approximation.
Indeed, $f$ is the parameter which governs the softness of the star-star as
well as of the star-colloid interaction, as we demonstrate below, and which
provides the natural bridge between the colloid-chain mixtures  (at $f = 1$ or
2) and the binary hard sphere mixture, formally equivalent to the
limit $f \rightarrow \infty$. Moreover, by changing $f$ or the degree of
polymerization $N$ of the star arms, a large domain of size ratios between the
components can be covered. 
When the stars are of comparable size with the
colloids a depletion picture of the mixture will be less accurate
than a full, two-component description because the former ignores the 
effects of many-body terms.
In this paper, we propose analytical expressions for the star-star and
star-colloid interactions for all $f$ values and for a large  range of size
ratios between the two. These are based on the one hand on accurate,
monomer-resolved computer simulations and on the other on theoretical arguments
regarding the functional form of these interactions. Using these expressions,
we make theoretical predictions on the mixing-demixing (or `gas-liquid')
transition in star polymer-colloid mixtures. Our measured demixing
curves are in good agreement with the theoretical predictions.

We consider a binary system with $N_{\rm c}$ colloidal spheres
of diameter $\sigma_{\rm c}$ (radius $R_{\rm c}$) and $N_{\rm s}$ star
polymers, characterized by a diameter of gyration $\sigma_{\rm g}$ (radius of
gyration $R_{\rm g}$) and an arm number $f$. For the special case $f=2$, star
polymers reduce to linear polymers and we reach the limit of colloid-polymer
mixtures 
\cite{ilett:etal:pre:95,lekkerkerker:etal:92,dijkstra:brader:evans:jpcm:99}. 
The total particle number is  $N=N_{\rm c}+N_{\rm s}$. Let $q=\sigma_{\rm
g}/\sigma_{\rm c}$ be the size ratio and  $\eta_{\rm c}=\frac{N_{\rm
c}}{V}\frac{\pi}{6}\sigma^{3}_{\rm c}$ and  $\eta_{\rm s}=\frac{N_{\rm
s}}{V}\frac{\pi}{6}\sigma^{3}_{\rm g}$ the packing fractions of the colloids
and stars respectively, enclosed in a volume $V$.  

Experimentally,  we studied two sets of star polymer-colloid mixtures
consisting of poly(methylmethacrylate) (PMMA) particles and poly(butadiene) 
(PB) star polymers  with size ratios $q\approx 0.49$ and $q\approx 0.18$,
respectively.  PMMA particles were synthesized following a standard procedure
\cite{antl:etal:86}.  Stock suspensions were prepared either in
cis-decahydronaphthalene  (cis-decalin) or cis-decalin/tetrahydronaphthalene
(tetralin) mixture as an index-matched solvent. These systems  have been
established as hard sphere models \cite{ilett:etal:pre:95}.  The volume
fraction $\eta_{\rm c}$ was calibrated using the onset of the hard sphere
freezing transition, taken to be at $\eta_{\rm c}=0.494$ and observed as the
nucleation of iridescent colloidal crystals.  The PB star polymers were
prepared by anionic polymerization  following an established procedure
\cite{allgaier:96,hadji:80}. Star arms were synthesized by polymerizing
butadiene with secondary butyl lithium as initiator.  The resulting living
polymer chains were coupled to the  chlorosilane linking agent having ideally
6, 16 and 32 Si-Cl-groups.  The molecular weights $M_{\rm w}$ of the PB arms
were adjusted to give star polymers with values of 
$\langle R_{\rm g}^2 \rangle^{1/2}=
0.0172M_{\rm w}^{0.609}f^{-0.403}$ \cite{Grest:review:96} as close
to $50\,{\rm nm}$ as possible.  A linear PB polymer ($f=2$) was prepared as a
reference system. The particles and star polymers were characterized using
light and small angle neutron  scattering (SANS) \cite{d11_ill}.  The results
are  summarized in Table \ref{tab:mol}. 

Samples were prepared by mixing PMMA suspensions with PB stock solutions. Each
sample was homogenized by prolonged tumbling and  allowed to equilibrate  and
observed by eye at room temperature $T=25\,^{\circ}{\rm C}$ \cite{joerg:unpub}.
In all samples with $q \approx 0.49$,  addition of polymer to suspensions with
$\eta_{\rm c} \sim 0.1-0.4$ brought about, successively, phase separation into
colloidal gas and liquid (or demixing), triple coexistence of gas, liquid and
crystal, and gas-crystal coexistence. In samples with $q \approx 0.18$,
addition of polymer first led to fluid-crystal coexistence; a metastable
gas-liquid binodal buried inside the equilibrium fluid-crystal coexistence
region was encountered at higher polymer concentrations \cite{poon:1995}. In
all cases, demixing started within several hours, crystallization within two
days. Here we focus on the demixing transition.

\begin{figure}
\vspace{5mm}
     \epsfxsize=6cm
     \epsfysize=4.5cm
     ~\hfill\epsfbox{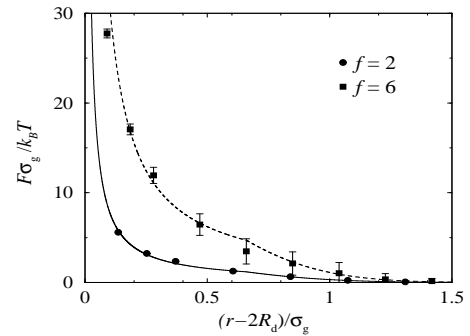}\hfill~
\caption{Effective force between two isolated star polymers for arm
  numbers $f=2,6$ at center-to-center distance $r$. The
  theoretical result (line) derived from
  eq.\ (\ref{pot_ss2}) is compared to computer simulation data
  (symbols) to obtain the decay parameter $\tau$.
  $R_{\rm d}$ is the nonvanishing core radius of one simulated star,
  values in Table \ref{tab2}. Error bars are shown for the case $f = 6$
  and provide an estimate for all $f$-values.}
\label{ss_force}
\end{figure}

Theoretically we model the thermodynamics of the mixtures on the
level of pair potentials between the two mesoscopic components, having
integrated out the monomer and solvent degrees of freedom. Thus, three
pair potentials are used as inputs for thermodynamically self-consistent
integral equations which are closed with the Rogers-Young (RY) scheme
\cite{rogers:young}. The colloid-colloid interaction at center-to-center
distance $r$ is taken to be that of hard spheres (HS): 
\begin{eqnarray}
V_{\rm cc}(r) = \cases {\infty & $ r \leq \sigma_{\rm c}$; \cr
   0 & else. \cr}
\label{pot_hs}
\end{eqnarray}
The effective interaction between two stars in a good solvent was
recently derived by theoretical scaling arguments and verified by
neutron scattering and molecular simulation, where the
monomers were explicitly resolved 
\cite{likos:etal:prl:98,jusufi:macromolecules:99}. The pair
potential is modeled by an 
ultrasoft interaction which is logarithmic for an inner core and
shows a Yukawa-type
exponential decay at larger 
distances \cite{likos:etal:prl:98,watzlawek:etal:prl:99}:
\begin{equation}
V_{\rm ss}(r)  = \frac{5}{18}k_{\rm B}Tf^{\frac{3}{2}} 
 \cases{
   -\ln(\frac{r}{\sigma_{\rm s}}) + \frac{1}{1+\sqrt{f}/2}
   & $ r \leq \sigma_{\rm s} $; \cr
   {\frac{\sigma_{\rm s}/r}{1+\sqrt{f}/2}
    \exp(-\frac{\sqrt{f}}{2\sigma_{\rm s}}(r-\sigma_{\rm s})) }
   & else,\cr} 
\label{pot_ss}
\end{equation}
with $k_{\rm B}T$ being the thermal energy. 
Our computer simulations show that the so-called
corona-diameter $\sigma_{\rm s}$ remains fixed for all considered  arm numbers
$f$, being related to the diameter of gyration through  $\sigma_{\rm
s}\simeq0.66\sigma_{\rm g}$ \cite{jusufi:macromolecules:99}. However,
the theoretical approach giving rise to 
eq.\ (\ref{pot_ss}) does not hold for arm numbers $f\lesssim 10$,  
because the
Daoud-Cotton model of a star \cite{daoud:cotton:82:1}, on which the Yukawa
decay rests, is not valid for small $f$. In these cases, the interaction
inclines to a shorter-ranged decay for $r > \sigma_{\rm s}$. This is
consistent with approaches in which at the linear polymer limit a  Gaussian
behavior of the pair potential is  predicted
\cite{ard:peter:00,bolhuis:preprint:00,krueger:etal:89}. Only the
large distance decay of the interaction is affected; its form 
at close approaches
has to remain  logarithmic \cite{Witten:Pincus:86:2}. Accordingly, we
propose the following star-star pair potential for arm numbers 
$f\lesssim 10$,
replacing the Yukawa by a Gaussian decay:
\begin{equation}
V_{\rm ss}(r)  = \frac{5}{18}k_{\rm B}Tf^{\frac{3}{2}} 
 \cases{
   -\ln(\frac{r}{\sigma_{\rm s}})+\frac{1}{2\tau^{2}\sigma_{\rm s}^{2}} & 
  $ r \leq \sigma_{\rm s} $; \cr
   {\frac{1}{2\tau^{2}\sigma_{\rm
 s}^{2}}\exp\left(-\tau^{2}(r^{2}-\sigma_{\rm s}^{2})\right) } & else,
 \cr}
\label{pot_ss2}
\end{equation} 
where $\tau(f)$ is a free parameter of the order of  $1/R_{\rm g}$ and is
obtained by fitting to computer simulation results, see 
Fig.\ \ref{ss_force} and Table \ref{tab2}. 
Using $\tau\sigma_{\rm s}(f = 2) = 1.03$ 
we obtain for the second virial coefficient of polymer solutions 
the value $B_2/R_{\rm g}^3 = 5.59$, in agreement with the estimate
$5.5 < B_2/R_{\rm g}^3 < 5.9$ from RG
and simulations \cite{bolhuis:preprint:00}.

\begin{figure}
\vspace{5mm}
     \epsfxsize=6cm
     \epsfysize=4.5cm 
     ~\hfill\epsfbox{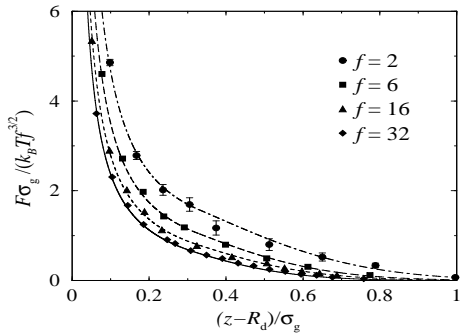}\hfill~
\caption{Effective force between an isolated star polymer and a hard
  flat wall $(q=0)$ for arm numbers $f=2,6,16$ and 32. 
  $z$ is the distance from
  the star center to the surface of the wall. Theoretical curves from 
  eq.\ (\ref{pot_sc}) were compared
  to computer simulation data (symbols) to obtain the prefactor
  $\Lambda$ and the decay parameter $\kappa$. For better comparison we
  divided the force by $f^{3/2}$.}
\label{sw_force}
\end{figure}

An analytic form for the star
polymer-colloid pair potential can be found by integrating the osmotic
pressure of one star along the spherical surface of a colloid, following an
idea put forward by Pincus \cite{pincus:porc:91}. 
This can be achieved for
arbitrary curvatures of the colloid but
the analytical
result below is 
accurate for size ratios $q\lesssim 0.7$ and reads \cite{jusufi:starcoll}:
\begin{eqnarray}
\nonumber
V_{\rm sc}(r) = \Lambda k_{\rm B}Tf^{\frac{3}{2}}
                  \frac{\sigma_{\rm c}}{2r}\times
\\
   \cases {\infty &  $ r<\frac{\sigma_{\rm c}}{2} $; \cr
   \xi_{2}-\ln(\frac{2z}{\sigma_{\rm s}})-(\frac{4z^{2}}
                 {\sigma_{\rm s}^{2}}-1)(\xi_{1}-\frac{1}{2})
       & $ \frac{\sigma_{\rm c}}{2}\leq r < 
              \frac{\sigma_{\rm s}+\sigma_{\rm c}}{2} $; \cr
   {\xi_{2}(1-{\rm erf}(2\kappa z))/(1-{\rm erf}(\kappa\sigma_{\rm s}))} 
   & else,\cr}
\label{pot_sc}
\end{eqnarray} 
where $z=r-\sigma_{\rm c}/2$ is the distance from the center of the
star polymer to the surface of the colloid. The constants are 
$\xi_{1}=1/(1+2\kappa^{2}\sigma_{\rm s}^{2})$ and
$\xi_{2}=\frac{\sqrt{\pi}\xi_{1}}
         {\kappa\sigma_{\rm s}}\exp(\kappa^{2}\sigma_{\rm s}^{2})
         (1-{\rm erf}(\kappa\sigma_{\rm s}))$.
$\Lambda(f)$ and $\kappa(f)$ are fit parameters, obtained from 
computer simulations where the force between an isolated star and a
hard flat wall is calculated, see Fig. \ref{sw_force}.
$\kappa$ is in order of $1/\sigma_{\rm g}$, see the values in
Table \ref{tab2}, whereas geometrical arguments yield a limit 
$\Lambda_{\infty} = 5/36$ for very large $f$.

To access the thermodynamics of the mixture, we solve the 
two-component Rogers-Young (RY) closure, which is reliable
for the one component star polymer system
\cite{watzlawek:etal:jpcm:98} and shows spinodal instability in highly
asymmetric hard sphere mixtures \cite{biben:hansen:prl:91}.  
We also performed standard Monte Carlo (MC)
simulations using the interactions (\ref{pot_hs})-(\ref{pot_sc}) as inputs and
measuring the structure factors at selected thermodynamics points. Excellent
agreement between RY and MC was found. The thermodynamic consistency of the RY
closure is enforced with a single adjustable parameter $\xi$;
a simple scaling of the form $\xi_{\alpha\beta}=\xi/\sigma_{\alpha\beta}, 
(\alpha,\beta={\rm c,s})$ showed only small differences compared to the
unscaled form. 

\begin{figure}
\vspace{5mm}
     \epsfxsize=6cm
     \epsfysize=4.5cm
     ~\hfill\epsfbox{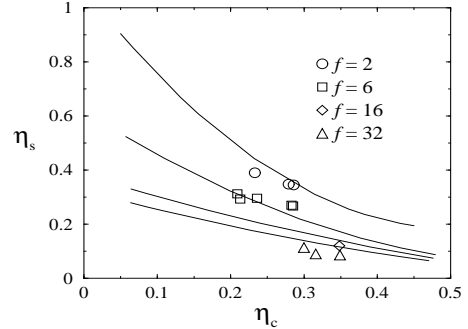}\hfill~
\caption{Binodals for the mixing-demixing transition in star
  polymer-colloid mixtures for different arm numbers $f=2,6,16,32$
  (from top to bottom)
  and size ratio $q\approx 0.49$. 
  Symbols mark experimental results compared with
  theory (lines) for $q=0.50$.}
\label{bino1}
\end{figure}

\begin{figure}
\vspace{5mm}
     \epsfxsize=6cm
     \epsfysize=4.5cm
     ~\hfill\epsfbox{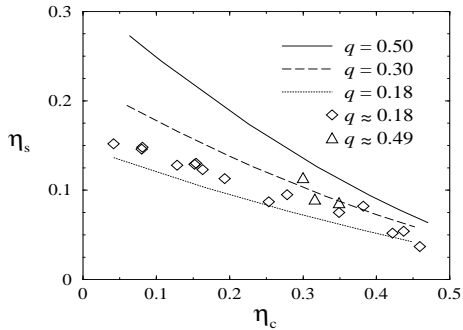}\hfill~
\caption{Same as in Fig.\ \ref{bino1}
         for an arm number $f=32$ and  different size
         ratios $q$.}
\label{bino2}
\end{figure}

The structure of binary mixtures is described by three partial static structure
factors $S_{\alpha\beta}(k)$, with $\alpha,\beta={\rm c,s}$ at wavevector $k$,
obtained from the integral equations. Indication of a demixing transition
is the divergence of all structure factors at the long wavelength limit
$k\rightarrow 0$, marking the {\it spinodal line} of the system. It is more
convenient to consider the concentration structure factor  
$S_{\rm con}(k)=x_{\rm s}^{2}S_{\rm cc}(k)+
                x_{\rm c}^{2}S_{\rm ss}(k)- 
               2x_{\rm c}x_{\rm s}S_{\rm cs}(k)$.
We denote the partial concentrations by $x_{i}=N_{i}/N,\; (i={\rm c,s})$.
The $k\rightarrow 0$ limit provides the approach to thermodynamics,
through the  relation \cite{biben:hansen:prl:91,bhatia:thornton:prb:70}: 
\begin{eqnarray} 
\lim_{k\rightarrow 0}S_{\rm con}(k)=
k_{\rm B}T\left[\frac{\partial^{2}g(x_{\rm c},P,T)}
{\partial x_{\rm c}^{2}}\right]^{-1},
\label{scon}
\end{eqnarray}
where $g(x_{\rm c},P,T)$ is the Gibbs free energy $G(x_{\rm c},N,P,T)$
per particle and $P$ denotes the pressure of the mixture. If the function
$g(x_{\rm c})$ has concave parts the system shows phase coexistence and the
phase boundaries can be calculated by the common tangent construction on the
$g(x_{\rm c})$ vs.\ $x_{\rm c}$ curves at constant $P$ and $T$,
ensuring the equality of the chemical potentials
for each component in the two phases. The results obtained are
shown in  Figs.\ \ref{bino1} and \ref{bino2}.

Inside the spinodal line, the limits $S_{\alpha,\beta}(k\to 0)$ attain
unphysical, negative values associated with the physical instability of the
mixture against phase separation. Consequently, a solution of the integral
equations is not possible there, and above the critical pressure $P^{*}$,
$S_{\rm con}(x_{\rm c},k=0)$ is unknown in some  interval $\Delta x_{\rm
c}(P)$. Thus, it is necessary to interpolate $S_{\rm con}(x_{\rm c},k=0)$  in
order to perform the integration of eq.\ (\ref{scon}). In the vicinity of the
critical point $\eta^{*}_{\rm c}\simeq 0.3$ the missing interval  $\Delta
x_{\rm c}$ is very small and the interpolation is reliable. Here the binodals
should be accurate, while for higher pressures (packing fractions  $\eta_{\rm
c}<\eta^{*}_{\rm c}$ and $\eta_{\rm c}>\eta^{*}_{\rm c}$) the binodals are more
approximate but  show reasonable behavior. For highly asymmetric systems
($q\lesssim 0.18$) it becomes more and more difficult to get solutions of the
integral equations in the vicinity of the spinodal line
and the calculation of binodals was not possible.

The results in Figs.\ \ref{bino1} and \ref{bino2} show that theory and
experiment are in good agreement. In particular, the same
trends are found as functions of the system parameters $f$ and $q$. By
increasing $f$  and keeping $q$ fixed, the demixing transition moves to lower
star packing fractions $\eta_{\rm s}$, as shown in Fig.\ \ref{bino1}. The
largest differences occur at low arm numbers $f\lesssim 10 $ which
is theoretcially caused by the major changes in the
star-colloid pair potential for low $f$. When $q$ is decreased but
$f$ remains fixed, again a motion of the  binodals to lower $\eta_{\rm s}$ is
observed, as seen in Fig.\ \ref{bino2}. This trend is {\it opposite}
to the one predicted by the AO model \cite{matthias}.
The crucial parameter that determines the trends of the phase
diagram is the non-additivity of the mixture.
Systems interpolating between the 
fully additive hard sphere mixture and the fully nonadditive AO model
can feature a depletion interaction which is more attractive
than in the AO limit \cite{roth:evans:epl}. The
agreement between theory and experiment is brought about {\it without} the use
of any free parameters in the former that would allow for a rescaling of  sizes
or densities. All values are read off from experiment and the only free
parameters of the theory appear on the level of the pair potentials and are
used only in order to fit analytical expressions to the
microscopically-determined star-star and star-colloid pair interactions.

To summarize, we have derived from first principles the effective
interactions between the components of a star polymer-colloid
mixture, proposing analytical expressions for these. 
We have used these to make theoretical predictions
about the thermodynamic behavior of the fluid phase of the system
finding very good agreement with the experimental results we 
obtained independently. Further work regarding, e.g., the 
crystallization properties of these mixtures is currently under way.     

We thank M.~Schmidt, A.~A.~Louis, R.~Finken and A.~Lang for useful
discussions, A.~Schlensog for technical support and the 
DFG for support within the SFB 237. 
J.S. is supported by the DFG and A.B.S. by NASA.

\begin{table}
\begin{center}
\begin{tabular}{c c c c c}
${\rm Monomer}$ & $f$ & $M_{\rm w} \cdot 10^{-6}$[g/Mol]$^{a)}$ &
$R_{\rm c} [{\rm nm}]^{b)}$ & $R_{\rm g}$[nm]$^{a)}$ \\
\hline
PMMA & -  & - & 104.0$\pm$ 2.5 & - \\
PMMA & -  & - & 289.0$\pm$ 4.5 & - \\
PB   & 2  & 0.57 $\pm$ 0.36 & - &51.0 $\pm$ 3.5 \\
PB   & 6  & 1.51 $\pm$ 0.06 & - &52.1 $\pm$ 0.6 \\
PB   & 16 & 3.45 $\pm$ 0.27 & - &51.1 $\pm$ 0.5 \\
PB   & 32 & 4.87 $\pm$ 0.39 & - &51.4 $\pm$ 0.5 \\
\end{tabular}
\end{center}
\footnotesize{$^{a)}$ small angle neutron scattering (SANS)
              \\
              $^{b)}$ static light scattering (SLS)}
\caption{Molecular characteristics of PMMA particles and PB star polymers.}
\label{tab:mol}
\end{table}

\begin{table}
\begin{center}
\begin{tabular}{c c c c c}
$\;\;f\;\;$ & $\;\;\Lambda(f)\;\;$ &
$\;\;\kappa\sigma_{\rm s}\;\;$ & $\;\;\tau\sigma_{\rm s}\;\;$ &
$\;\;R_{\rm d}/\sigma_{\rm g}\;\;$  \\
\hline
2  & 0.46 & 0.58 & 1.03 & 0.04 \\
6  & 0.35 & 0.71 & 1.16 & 0.03 \\
16 & 0.28 & 0.76 & - & 0.04 \\
32  & 0.24 & 0.84 & - & 0.06 \\
\end{tabular}
\caption{Fit parameters $\Lambda,\kappa,\tau$ for the effective
  star-wall interaction of eq.\ (\ref{pot_sc}) and the star-star interaction
  eq.\ (\ref{pot_ss2}) obtained from molecular simulation. $R_{\rm d}$
  is the nonvanishing core radius of one simulated star.}
\label{tab2}
\end{center}
\end{table}

\end{document}